\documentclass[aps,preprint,groupedaddress]{revtex4-1}

\usepackage{graphicx}
\usepackage{color}
\usepackage{comment}
\usepackage{subfig}
\newcommand{\minitab}[2][l]{\begin{tabular}{#1}#2\end{tabular}}

\begin{document}


\title{Emoticon-based Ambivalent Expression: \\A Hidden Indicator for Unusual Behaviors in Weibo}


\author{Yue Hu$^1$, Jichang Zhao$^{2,~1}$ and Junjie Wu$^1$}
\affiliation{
$^1$School of Economics and Management, Beihang University \\
$^2$State Key Lab of Software Development Environment, Beihang University\\
}


\begin{abstract}
Recent decades have witnessed online social media being a big-data window for quantificationally testifying conventional social theories and exploring much detailed human behavioral patterns. In this paper, by tracing the emoticon use in Weibo, a group of hidden ``ambivalent users'' are disclosed for frequently posting ambivalent tweets containing both positive and negative emotions. Further investigation reveals that this ambivalent expression could be a novel indicator of many unusual social behaviors. For instance, ambivalent users with the female as the majority like to make a sound in midnights or at weekends. They mention their close friends frequently in ambivalent tweets, which attract more replies and thus serve as a more private communication way. Ambivalent users also respond differently to public affairs from others and demonstrate more interests in entertainment and sports events. Moreover, the sentiment shift of words adopted in ambivalent tweets is more evident than usual and exhibits a clear ``negative to positive'' pattern. The above observations, though being promiscuous seemingly, actually point to the self regulation of negative mood in Weibo, which could find its base from the emotion management theories in sociology but makes an interesting extension to the online environment. Finally, as an interesting corollary, ambivalent users are found connected with compulsive buyers and turn out to be perfect targets for online marketing.
\end{abstract}

\keywords{Ambivalent expression, Ambivalent users, Self-regulation, Weibo}

\maketitle

\section{Introduction}
\label{sub:intro}

Emotion expression is a psychological behavior with the purpose of communicating affective states between different individuals. This behavior could be either verbal or nonverbal, including text, voices, faces and bodies~\cite{valenza-2013}. In the epoch of the Internet, tremendous developments of online social media provide abundant innovative and powerful means of information exchange, bringing unprecedented richness and diversity to the ways of emotional expression. 

Among these new forms, \emph{emoticon} is getting more and more popular with the rapid growth of Weibo, a Twitter-like service in China. In general, an emoticon is a kind of non-verbal sentiment language represented by a vivid image, which differs itself from a \emph{smiley} formed by combining punctuation marks as used in Twitter. Indeed, users seem to show increasing interests in expressing sentiments through around 2,000 emoticons designed by Weibo, ranging from facial expressions like laughing or bursting into tears, body languages like applause or hug, to some icons like the sun, Christmas trees or birthday cakes. This is not unusual, since tweets are very short and context-absent for the 140-word limitation, while emoticons could easily convey specific emotions in a vivid and personalized way.

It has been found that both smileys and emoticons are strongly related with typical sentiment words, and could serve as convincing indicators of different emotions~\cite{Liu10sentimentanalysis}. Tossell et al. confirm that emoticon usage is contextual, and people use more negative emoticons than positive ones~\cite{Tossell}. Researchers also realize that emoticons and smileys could be effective features for texts to improve the precision of sentiment analysis~\cite{emoticon_review,moodlens,eacl2012,liuaaai12,usa_huxia,cikm2013}, or be treated as sentiment labels to avoid intensive labor costs for preparing training samples~\cite{moodlens}. 

As a speical social behavior, emoticon usage is shaped by cultural and social factors. The pattern of emoticon usage in short-message texting is investigated and the discrepancies between male and female users are revealed~\cite{Derks}. For instance, female users send more messages containing emoticons, while male users use a wider range of emoticons. Schnoebelen et al. distinguish users by whether they use nose smiley ``:-)'' or non-nose smiley ``:)'', and demonstrate that the variants correspond to different types of users, tweeting with different vocabularies and writing styles~\cite{smiley_nose}. Park et al. point out that emoticons are socio-cultural norms and their meanings could be affected by the identity of the speaker~\cite{park_icwsm2013}. 

The above research, though being very interesting, is mainly focused on emoticons rather than emoticon users. Who they are, what they are talking about, how they behave online and why they adopt emoticons--- these human-centric problems are actually very interesting to sociologists and perhaps marketers, who are always seeking for potential consumers.Meanwhile, different from expensive but spatio-temporally limited surveys in tradition~\cite{wood-2013}, digital traces created by interactions with technology indeed offers a new probe to collective human behavior and these new data are fuelling the rapid development of computational social science~\cite{preis-2013}. These indeed motivates our study in this paper, which takes ambivalent emoticons in a tweet as latent clue for tracing the unusual behaviors of a speical group called ``ambivalent users''. 

More specifically, we manually select 79 positive and 36 negative emoticons with unambiguous polarities to label the sentiments of tweets; that is, tweets containing only positive (negative) emoticons are defined as positive (negative) tweets. Surprisingly, we find many \emph{ambivalent tweets} containing both positive and negative emoticons, indicating inconsistent emotions~\cite{consistent_theory}. The \emph{ambivalent users} are then defined as the users who have published more than 30 ambivalent tweets in 2012. The \emph{senior ambivalent users} are defined accordingly as a group of special ambivalent users who have published more than 50 ambivalent tweets in 2012. As contrast, the \emph{ordinary users} are the users who have never published any ambivalent tweet in 2012. After excluding abnormal users like verified celebrities, organizations, inactive users (post less than 12 tweets in 2012 or have less than 65 followers) and spam accounts, we finally have 1,069 ambivalent users, among whom 357 are senior users, and the ordinary users total 46,245. 

A first study of demographics shows that ambivalent expression is more popular in Weibo than in Twitter, and the female occupies a significantly higher proportion of ambivalent users. This indeed validates the conventional theory in psychology that people from east show significantly stronger association of positive and negative in emotion expression than the west, especially the female~\cite{association_theory}. Further explorations on the behaviors of ambivalent users disclose more interesting patterns. Specifically, ambivalent users like to express ambivalence in midnights or at weekends, focus much more on topics of entertainment and sports, actively mention their close friends in tweets and gain more replies rather than reposts than ordinary users, and frequently use positive terms while expressing negative mood. 

The above observations suggest the subtle link between users' mood status and the usage of online social media, which has become a very hot topic in recent years and attracted increasing research interests~\cite{happiness_assortative,dodds-2011,happiness_correlation,lewis-2013,frank-2013,fan2014anger}. Specifically, based on the theoreies of emotion management in sociology, we point out that ambivalent users' unusual behaviors can be well explained from the view of \emph{self regulation} against negative feelings. Moreover, this self-regulation behavior of ambivalent users seems to be conscious, which makes our finding distinct from existing studies mostly on unconscious behaviors. Finally, as a natural conjecture, we testify the self-correcting effect of shopping on ambivalent users' negative mood using the tweets posted around the Singles' Day in 2012. This indicates that ambivalent users like compulsive users are ideal targets for online marketing.

\section{Results}
\label{sub:res}

\subsection{Ambivalent Expression}
We first observe ambivalent expression on Weibo. It is interesting that nearly 1.9\% of the emotional tweets on Weibo are ambivalent, but it reduces to 1.1\% for Twitter, indicating stronger inclination of Chinese Weibo users for ambivanent expression (see \emph{Methods}). This may owe to the extraordinary vividness and ease-of-use of the built-in emoticon modular of Weibo, and the linguistic differences between Chinese and English may also contribute.

We then focus on the ambivalent users on Weibo. It is interesting that ambivalent users generally showed more passive mood by publishing significantly more negative tweets than ordinary users (the negative-tweet ratios are 27.1\% against 18.8\% for ambivalent and ordinary users, respectively, in 2012). Moreover, the female occupies 79.0\% of total ambivalent users, but only 42.8\% of the ordinary users. Among them we find a verified community named {\em Weibo Lady}, whose members are most active females passionate about sharing their personal lives. In this community, 82.8\% of the users ever posted ambivalent tweets, which is consistent with the theory that people from the east show a stronger association between positive and negative affect than the west, especially the female~\cite{association_theory}. 

\begin{figure}[ht]
\centering
\includegraphics[scale=0.5]{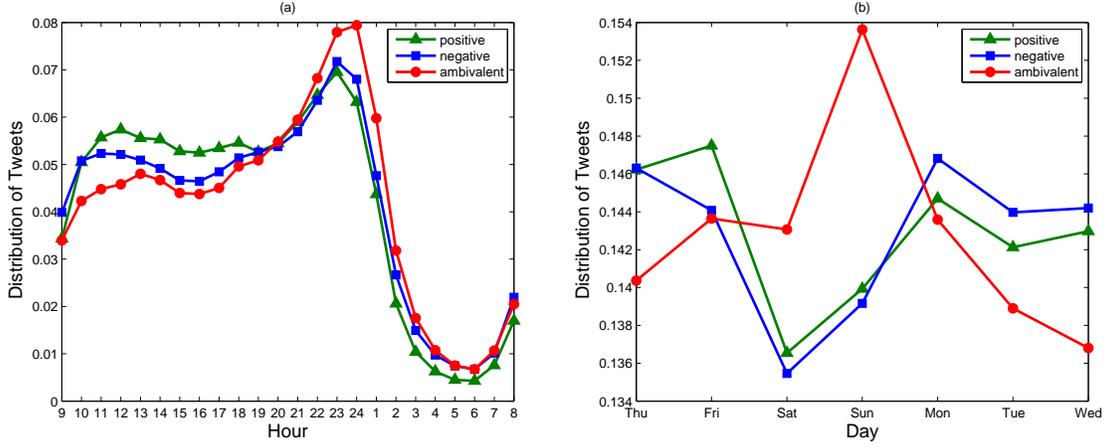}
\caption{Distribution of tweets posted at different time. }
\label{fig:tweet_time_pattern}
\end{figure}

We further observe the posting behavior of ambivalent users. Fig.~\ref{fig:tweet_time_pattern}(a) shows the daily pattern, where ambivalent tweets seem more noticeable in the midnight from 11PM to 2AM. Similar situation happens for the weekly pattern in Fig.~\ref{fig:tweet_time_pattern}(b), where significantly more ambivalent tweets are posted at the weekend, especially on Sunday. Since midnights and weekends are usually the leisure time for individuals, we could conjecture that ambivalent users are apt to express mixed feelings when relaxed.

\subsection{Topic Preference}

\begin{figure}
\centering
\includegraphics[scale=0.5]{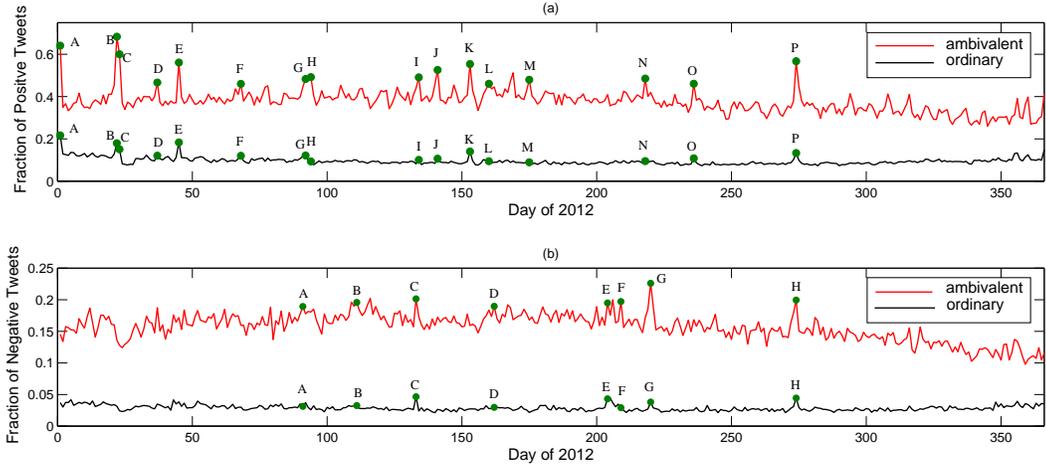}
\caption{Comparison of topic preferences between ambivalent and ordinary users. (a) Variation of positive emotion for topic detection. (b) Variation of negative emotion for topic detection.}
\label{fig:events}
\end{figure}

We further explore the topic preference of ambivalent users. It has been reported that the variation of social behavioral indexes, like ``the number of posts'', could be employed to detect hot topics or external events in social media~\cite{moodlens,tweet_rate}. Accordingly, we trace the time series of positive and negative tweets, respectively, and highlight the peaks with clear event semantics (see {\em Methods}), as shown in Fig.~\ref{fig:events}.

Fig.~\ref{fig:events}(a) suggests that, in terms of positive emotion, besides the festivals such as Spring Festival, Valentine's Day, and Moon Cake Festival, ambivalent users enjoy talking more about some niche topics of sports and entertainment, like Day to speak out love ({\em J}), a singer, Xin Liu's Birthday ({\em L}), European Cup competitions ({\em L} and {\em M}), and Olympic Games ({\em N}). The variations of negative sentiment in Fig.~\ref{fig:events}(b) further validate this point; that is, besides the natural disasters with universal concerns, ambivalent users pay more attention to topics like the rescue of dogs ({\em B}), the competitions in European Cup ({\em D}), the final episode of a TV play ({\em F}) and the final of Voice of China ({\em H}). These unveil that ambivalent users are indeed more sensitive and emotional than the ordinary ones. The extracted topic words in Tables~\ref{tab:pos_event} and \ref{tab:neg_event} provide more details (see {\em Methods}).

\begin{figure}
\centering
\includegraphics[scale=0.7]{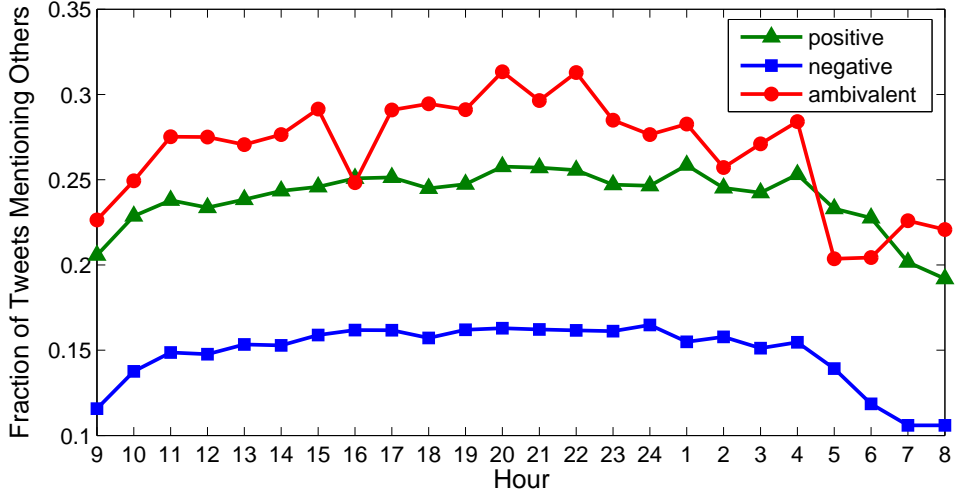}
\caption{Fraction of tweets mentioning others at different hours.}
\label{fig:mention_pattern}
\end{figure}

\subsection{Social Interaction}
It is interesting that ambivalent tweet is an important channel for ambivalent users to socialize online. To illustrate, we take senior ambivalent users as well as ordinary users for comparative study. Simple statistics show that ambivalent users are a group of people who are socially more active online, e.g., having more fans (2506 versus 735) and bi-friends (270 versus 194) and posting more tweets (1890 versus 336), than the ordinary users.

More interestingly, ambivalent users have a subtle communication mode. To illustrate this, we trace the ``@'' behavior on all the tweets they posted in 2012 (a Weibo user can use the ``@'' symbol in her tweet to inform someone deliberately, which could be regarded as a way for more ``private'' communication on Weibo). Fig.~\ref{fig:mention_pattern} shows that ambivalent users are generally more apt to mention a friend in an ambivalent tweet. Further exploration reveals that the friends mentioned in ambivalent tweets share averagely 20 common followers with the ambivalent users, but the number reduces to 12 for the friends mentioned in positive tweets and even 7 in negative tweets. Since to share more friends indicates a stronger social tie between two users~\cite{tie_strength}, the above implies that ambivalent users prefer to interact with their closer friends via ambivalent tweets. In other words, we can touch close ties on Weibo by simply checking who is mentioned by whom in ambivalent tweets.

The reactions from the friends of ambivalent users are also very interesting. To illustrate this, we observe the \emph{replies} and \emph{reposts} of the tweets posted by senior ambivalent users in 2012. We find that ambivalent tweets are generally more apt to attract \emph{replies} --- a tweet containing dialectic emotions has 36 replies on average, whereas a positive (negative) tweet only has 34 (23) replies. This contradicts with the \emph{reposts'} case, where a positive tweet elicits more reposts than an ambivalent tweet (63 versus 53). Since a reply is a more private feedback than a repost, the above implies that ambivalent users are easier to gain private feedbacks via ambivalent tweets. In a nutshell, we can conclude that ambivalent users on Weibo depend heavily on ambivalent tweets to interact with their close friends in a relatively private way.

\subsection{Sentiment Shift}
We here demonstrate that posting ambivalent tweets can lead to sentiment shift and thus help to ease negative emotion. This is particularly important for ambivalent users, who seem generally more passive than ordinary users. This may also uncover the root of ambivalent expression on social media. 

To this end, we collect all the ambivalent tweets posted in 2012 and analyze the sentiment shift indicated by ambivalent emoticons within a tweet. Simple statistics reveal that there are 60,293 tweets shifting from negative emotion to positive one, but only 39,457 in reverse (we abandon tweets with ambiguous sentiment shift). The ``negative $\rightarrow$ positive'' pattern of emotional shift implies that Weibo indeed can ease negative mental state of ambivalent users.

We also try ``perceiving'' subtle emotional shifts from the words adopted in ambivalent tweets. To this end, we collect all the ambivalent tweets posted in 2012 and split them into short clauses, which are then tagged as positive (negative) clauses if they only contain positive (negative) emoticons. The frequent terms are then extracted from the positive (negative) clauses, and are compared with the terms used in purely positive (negative) tweets (see {\em Methods}). It is interesting that the sentiment shift in ambivalent tweets in the term level is surprisingly evident. Specifically, the ambivalent users tend to use some special terms in negative (positive) clauses that are actually more frequently occurred in positive (negative) tweets. This contradiction is particularly evident for the case of negative clauses, implying the more positive inclination of ambivalent tweets. This agrees with the mainstream pattern of emotion shift: negative $\rightarrow$ positive, and implies that posting ambivalent tweets indeed can help Weibo users realize self-correcting. We will have a more detailed discussion below.

\section{Discussion}
\label{sub:dis}

The sentiment shift of ambivalent tweets gives an important clue that online social media might affect or even manipulate users' mood status, which indeed has attracted specialized interests from researchers of various domains. For example, it has been found that socialising, information-seeking, and entertainment in online social media significantly influence user's positive emotion~\cite{Vanessa_positive_mood}, persistent emotional expressions for individual users and channels are revealed in online chat rooms~\cite{garas-2013}, and positive terms in news feed may trigger online friends to express similar feelings~\cite{facebook_emotion_spread}. From the dark side, however, disordered online social networking use leads to difficulties with emotion regulation~\cite{disorder_use}, and daily time spent on social networking is related to depression~\cite{depression}. These pilot studies, though being very interesting, have not touched the speical group of ambivalent users hidden inside online social media. Nor did they provide adequate explanations to the abnormal behavioral patterns of these users. In particular, the sentiment shift from negative to positive implies that ambivalent users ``actively'' leverage the Weibo platform to realize self-correcting, which has clear contrast to the ``unconscious'' emotional influences discussed in the above-mentioned literature.

Nevertheless, we can find some evidence for the self-regulation of negative emotions from earlier sociological research. Indeed, based on offline survey studies, these research reveals that people could tune their moods through music, social interactions, enjoyable activities, shopping, religion, distraction~\cite{people-religion} and cognitive reappraisal~\cite{cognition}. These results, though being widely accepted in the field of sociology, have hardly been tested in an online environment. Therefore, it would be very interesting if our findings on ambivalent users in Weibo could resonate with these offline findings. Indeed, as illustrated below, we can really explain the complicated behavioral patterns of ambivalent users from a sentiment self-regulation perspective.

First, as can be found in Sect. II.A, ambivalent users are apt to express passive moods and post ambivalent tweets in the midnights or at the weekends. This justifies that ambivalent users objectively have the need for self regulation of negative emotions, espeically when they are not occupied by work. Hence, we can conjecture that publishing tweets should be an ideal way to distract them from negative mentality, especially the ambivalent tweets that can help them realize cognitive reappraisal. To understand the last point, recall the finding in Sect. II.C that ambivalent users demonstrate an unusual sentiment-shift behavior, most evidently from negative mood to positive mood. This pattern implies that ambivalent users try to reduce their passive feelings through building new cognitions to the context, including opinions, attitudes, sentiments, etc.

Second, as illustrated in Sect. II.B, ambivalent users are particularly fond of topics like entertainment and sports. These enjoyable activities could arouse relaxed feelings and thus might liberate them from depression or stress. Moreover, the special social interaction patterns of ambivalent users given in Sect. II.C. further hint the existence of self regulation. That is, ambivalent users prefer to mention their close friends in tweets and get encouraging feedbacks from those friends. This coincides with previous findings from surveys that supportive interactions with friends produce positive affect~\cite{Oh_life_sastisfaction} and positive feedbacks on the profile enhance social self-esteem and well-being in online social networking~\cite{positive_feedback}. More interestingly, our findings suggest that interacting with close friends through more private actions like ``@'' and ``reply'' in online social networking can help boost self regulation. This subtle detail exhibits the possible uniqueness of self regulation theory in an online environment, which cannot been touched in traditional offline studies. 

\begin{figure}
\centering
\includegraphics[scale=0.6]{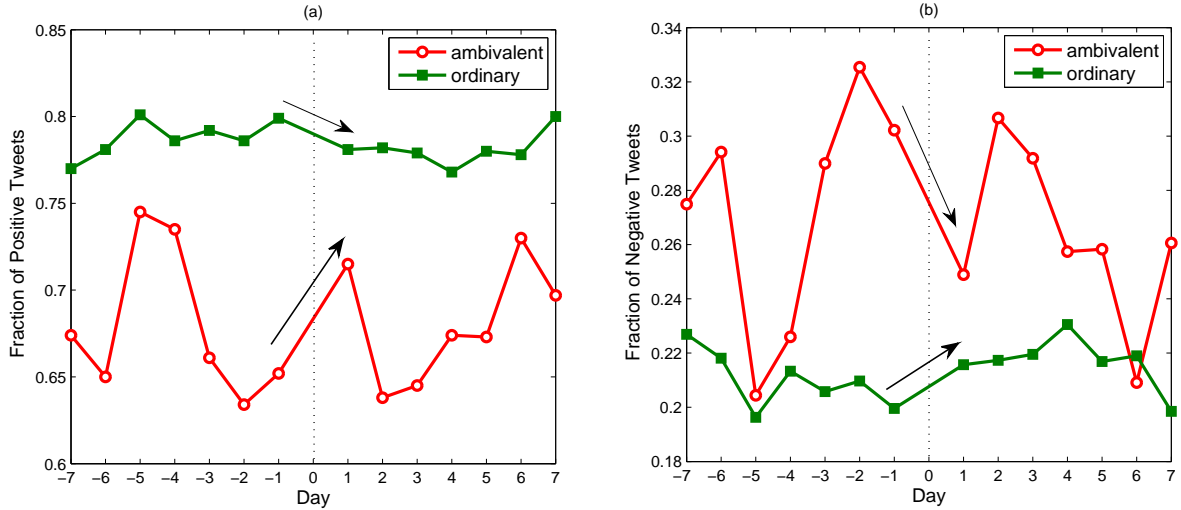}
\caption{Fractions of tweets posted in three days before and after shopping. For $x$ axis, the unit is a day and 0 stands for the shopping date. (a) Positive tweets. (b) Negative tweets.}
\label{fig:shopping_tweet_time_pattern}
\end{figure}

In a nutshell, the self regulation theory in sociology indeed well explain the distinct behaviors of ambivalent users in Weibo, and social media big data even enable us to find more distinct details in an online environment. It is also noteworthy that the self regulation of ambivalent users seems to be conscious, which makes our finding an important supplement to existing online studies. Along this way, one interesting corollary according to the self regulation theory is that ambivalent users would like to use shopping to self-correct negative mentality. We test this point by finding the shopping behaviors of ambivalent users around the Singles' day (or Doule Eleven, a famous day for promotion sales in China advocated by Taobao inc.) in 2012 and observing their emotional shifts (see \emph{Methods}). It is interesting that similar to compulsive buyers~\cite{compulstive_buyers}, ambivalent users experience an evident increment in positive mood after shopping. Specifically, around 20.5\% of ambivalent users post shopping tweets during the promotion sales while the ratio for ordinary users is just 7.8\%. Moreover, as shown in Fig.\ref{fig:shopping_tweet_time_pattern}, in time window of one day before and after shopping, the fraction of positive tweets posted by ambivalent users grows from 65.2\% to 71.5\% (negative tweets drop from 30.2\% to 24.9\%), while the value for ordinary users seems dropped slightly. This reversed shift of sentiments suggests that ambivalent users indeed regulate their negative feelings actively through shopping, which makes them ideal targets for online marketing.

\section{Conclusion}
\label{sec:con}

In this paper, we take ambivalent expression in Weibo as an important clue for tracing unusual online behaviors of ambivalent users. Our findings first confirm traditional sociological studies by showing that the ambivalent expression is preferred by the east, especially the female. Further investigation reveal various interesting behaviors of ambivalent users, including the topic preference in entertainment and sports events, the increasing communication desire when relaxed, and the private contact to close friends via @ symbol in ambivalent tweets as well as the replies as feedback. Moreover, we find a clear negative to positive mode in sentiment shift in ambivalent tweets, which guides us to find reasonable explanations to the above behaviors via the conventional theory of emotion management in sociology. As a result, we conclude that ambivalent expression is an effective way to self-regulate negative feelings in online social media, which is an important supplement to traditional offline sociological studies. Extended studies on the self-correcting effect of shopping suggest that our findings could shed light on applications like online marketing.

This study inevitably has limitations. First, for the online users investigated, the demographics other than gender, like age, occupation and geo-distribution,  are not well discussed for data limitation, and their correlations with ambivalence expression thus remain unclear. Second, because of missing a complete online following graph, this study is not embedded into the context of a social network, which otherwise might help us structurally characterize the detailed mechanism of emotional interplay between different users. Therefore, further explorations focusing on the above points would be interesting directions in future work. 

\section{Methods}
\label{sub:med}

{\em Weibo Data.}~Weibo is a Twitter-like service in China, which has accumulated more than 500 million registered users since founded in 2009. We select 137,981 users by breadth-first search starting from several Weibo-verified seeds and then crawl each user's profile page. We collect totally 68 million tweets posted by these users in 2012, among which 22.7\% are emoticonized tweets. In this paper, we only conduct the experiment on the data set of 2012, because from April of 2013, Alibaba has been a strategic cooperative partner of Weibo and Weibo therefore experiences a sudden growth in marketing tweets, which might badly contaminate the analysis. We select emoticons appeared more than 1,000 times and manually label them as positive or negative according to their images and descriptive words. To ensure the consistency of an emoticon's sentiment orientation, we check the top-200 frequently used keywords extracted from the 1,000 tweets containing that emoticon. The results show that many popular emoticons are strongly and consistently bonded with one specific emotion. Finally, we have 115 verified emoticons, among which 79 are positive and 36 are negative. They appear in more than 84.3\% of the emoticonized tweets, which consist of 312,456 ambivalent tweets, 13,430,096 positive tweets and still 2,858,413 negative ones. Our data sets can be freely downloaded from \url{www.datatang.com/data/47207} or \url{http://pan.baidu.com/s/1mg67cbm}.

{\em Twitter Data.}~For the comparison purpose, 467 million Twitter posts published by 20 million users are employed in this study~\cite{twitter_dataset}, covering about 30\% of all public tweets posted in a 7-month period from June 1, 2009 to Dec. 31, 2009. We use 15 positive smileys (i.e., ``:)", ``:D", ``=D'', ``=)'', ``:]'', ``=]'', ``:-)'', ``:-D'', ``:-]'', ``;)", ``;D'', ``;]'', ``;-)'', ``;-D'' and ``;-]'') and 9 negative smileys (i.e., ``:('', ``=('', ``:['', ``=['', ``:-('', ``:-['', ``:’('', ``:’['' and``D:''), which are also adopted in~\cite{smiley_list} for polarity classification. Finally, we find 34 million tweets containing at least one smiley, among which 28 million are positive, 6 million are negative, and only 385,145 are ambivalent.

{\em Computation of Topic Preference.}~We draw everyday tweets published in 2012 by ambivalent and ordinary users, respectively, for topic inference. Let $p_u^e(t)$ be the number of tweets published by user $u$ at day $t$ with emotion $e$ (either positive or negative). Let $\bar{p}_u$ be the average number of tweets published daily by $u$. Then for a group of users $G$, the emotional intensity of $e$ at day $t$ is given by $p_G^e(t)=\sum_{u\in G}{p_u^{e}(t)/\bar{p}_u}$. We trace $p_G^e(t)$ continuously along the year 2012, with $G$ being the group of ambivalent and ordinary users and $e$ being the positive and negative emotions, respectively, to find emotional peaks that might indicate real events. We finally obtain 16 and 8 peaks with clear event semantics from the top-20 peaks in positive and negative emotion lines, respectively, as shown in Fig.~\ref{fig:events}. The topics discussed by ambivalent users with extracted frequent terms are reported in Table~\ref{tab:pos_event} and Table~\ref{tab:neg_event}.

\begin{table}\scriptsize
\caption{Positive Topics Discussed by Ambivalent Users \label{tab:pos_event}}
\begin{tabular}{l}
\hline
{\bf A: Jan 1, 2012, New Year} \\[-0.1cm]
New Year, happy, one year, the first day, hope, ha-ha, 2011, happiness, together \\
{\bf B: Jan 22, 2012, Eve of lunar New Year} \\[-0.1cm]
{\minitab[l]{New year, happy, Spring Festival Gala, Year of Dragon, ha-ha, one year,\\[-0.1cm] happiness, celebrate the spring festival, healthy, friends}} \\
{\bf C: Jan 23, 2012, Spring Festival} \\[-0.1cm]
New year, happy, Year of Dragon, ha-ha, healthy, one year, happiness, first \\
{\bf D: Feb 6, 2012, Lantern Festival} \\[-0.1cm]
Happy, Lantern Festival, rice dumpling, happiness, firework, Valentine's Day of China \\
{\bf E: Feb 14, 2012, Valentine's Day} \\[-0.1cm]
Valentine's Day, happy, together, gifts, lovers, chocolate, enjoy, husband, dear, bachelor\\
{\bf F: Mar 8, 2012, Women's Day}\\[-0.1cm]
Happy, festival, Mother, thanks, women, enjoy, March 8th, girls \\
{\bf G: Apr 1, 2012, April Fools' Day}\\[-0.1cm]
Ha-ha, happy, April Fools Day\\
{\bf H: Apr 3, 2012, Weibo opened comments to the users}\\[-0.1cm]
Comments, Weibo, finally, the first comment\\
{\bf I: May 13, 2012, Mothers' Day}\\[-0.1cm]
Mother, happy, Mothers' Day, happiness, healthy, I love you, enjoy\\
{\bf J: May 20, 2012, Day to speak out love}\\[-0.1cm]
Ha-ha, happy, love, I love you, happiness, thanks, fighting, smile, speak out of love \\
{\bf K: Jun 1, 2012, Children's Day}\\[-0.1cm]
Children's Day, care, I love you, ha-ha, festival, enjoy, gifts, kids, lovely, cute\\
{\bf L: Jun 8, 2012, Xin Liu's Birthday and the beginning of European Cup}\\[-0.1cm]
Birthday, happy, Xin Liu, thanks; European Cup, beginning\\
{\bf M: Jun 23, 2012, Dragon Boat Festival and competitions of European Cup}\\[-0.1cm]
Dragon Boat Festival, zongzi, Germany, fighting, Greece, European Cup\\
{\bf N: Aug 5, 2012, Dan Lin won the gold medal}\\[-0.1cm]
{\minitab[l]{Fighting, Dan Lin, China, Super Dan, Olympic Games, Chong Wei Lee,\\[-0.1cm] Yang Sun, congratulations, Champaign, badminton}}\\
{\bf O: Aug 23, 2012, Star Festival}\\[-0.1cm]
Star Festival, happy, ha-ha, Lovers' Day, love, smile, together, gifts\\
{\bf P: Sep 30, 2012, Moon Cake Festival and the finals of the Voice of China}\\[-0.1cm]
{\minitab[l]{Mookcake Festival, Voice of China, mooncake, birthday of China, go home,\\[-0.1cm]
together, Mochou Wu, National Day, moon, Bo Liang}}\\
\hline
\end{tabular}
\end{table}

\begin{table}\scriptsize
\caption{Nagative Topics Discussed by Ambivalent Users \label{tab:neg_event}}
\begin{tabular}{l}
\hline
{\bf A: Mar 31,2012, Sina closed the comments in Weibo}\\[-0.1cm]
Comments, reposts, mean jokes, Sina, rumor, close\\
{\bf B: Apr 20, 2012, Volunteers rescued dogs in Kunming}\\[-0.1cm]
Dogs, raining, rainstorm, Guangzhou, animals, dog dealers, volunteers, Kunming\\
{\bf C: May 12, 2012, The 4th Anniversary of Wenchuan Earthquake}\\[-0.1cm]
{\minitab[l]{Four years, Wenchuan, victims, rest in peace, 5.12, earthquake, survivals,\\[-0.1cm]
fellows, blessing, commemorate}}\\
{\bf D: Jun 10, 2012, Group Stages in European Cup}\\[-0.1cm]
Holland, European Cup, fighting, Italy, Germany, Portugal, Denmark\\
{\bf E: Jul 22, 2012, Torrential Rains in Beijing}\\[-0.1cm]
{\minitab[l]{Beijing, rest in peace, salute, heroes, rainstorm, sacrifice, death,\\[-0.1cm]
local police station, policemen}}\\
{\bf F: Jul 27, 2012, Witness Insecurity}\\[-0.1cm]
Ending, Witness Insecurity, Mr Xu, TVB, die\\
{\bf G: Aug 7, 2012, Xiang Liu Got Injured in the Olympic Games}\\[-0.1cm]
Xiang Liu, China, London, Olympic Games, cry, fail down, referee, hero, fighting\\
{\bf H: Sep 30, 2012, The final of Voice of China}\\[-0.1cm]
Voice of China, Mochou Wu, ads, Zhiwen Jin, shady deal, Bo Liang, Sherry Chang Huei-mei\\
\hline
\end{tabular}
\end{table}

{\em Computation of Emotional Shift.}~We take all the emoticonized tweets containing at least one of the 115 verified emoticons to study the emotional shift inside a clause of ambivaent tweets. We first cut ambivalent tweets into short clauses by punctuation characters or whitespaces, and label the clauses containing only positive emoticons as positive clauses. We then extract the top-2,000 frequent keywords from the positive clauses, denoted as $W_{pa}$. Similarly, we extract top-2,000 keywords from purely positive tweets and form the keyword set $W_{pp}$. The different set $\tilde{W}_p=W_{pa}\setminus W_{pp}$ with $|\tilde{W}_p|=432$ then indicates the distinctive keywords used in ambivalent tweets to express positive feeling, which are subject to the emotional orientation test below. That is, for any keyword $t\in\tilde{W}_p$, we compute the occurrence frequencies of $t$ in postive and negative tweets, denoted as $f_p(t)$ and $f_n(t)$, respectively. Let $f_p(\tilde{W}_p)=\sum_{t\in\tilde{W}_p} f_p(t)/|\tilde{W}_p|$ denote the postive intensity of $\tilde{W}_p$ and $f_n(\tilde{W}_p)=\sum_{t\in\tilde{W}_p} f_n(t)/|\tilde{W}_p|$ denote the negative intensity. It is interesting to find that $f_p(\tilde{W}_p):f_n(\tilde{W}_p)=1.72\%:3.23\%$, indicating that the positive clauses of ambivalent tweets actually convey more negative feelings than purely positive tweets. We also find this subtle mismatch from the negative clauses of ambivalent tweets. Let $\tilde{W}_n$ denote the distinctive keywords used in ambivalent tweets to express negative feelings with $|\tilde{W}_n|=423$, we have $f_p(\tilde{W}_n):f_n(\tilde{W}_n)=4.17\%:1.67\%$, indicating the stronger positive feelings in ambivalent tweets' negative clauses.

{\em Computation of Shopping Behaviors.}~We collect all the tweets published by ambivalent users from Nov. 4 to Nov. 18, 2012 for identifying their shopping behaviors around the Singles' Day. The tweets containing keywords like ``buy'', ``shopping'' and ``taobao'' are then selected as candidates and subject to manual labeling for finding shopping tweets depicting users' real shopping behaviors. For each shopping tweet, we trace back to its author's tweeting history and calculate the fraction of positive and negative tweets, respectively, in the specified time window before and after shopping. The results of corresponding ambivalent or ordinary users are then averaged to get the final shift patterns. Note that we here neglect neutral tweets without emoticons. Statistical significance of sentiment shift is also testified by randomly shuffling the posting time of each user's tweets, after which the sentiment shift produced by the shopping behavior disappears.

\begin{acknowledgments}
This work was partially supported by National Natural Science Foundation of China (71322104,
71171007, 71471009), National Center for International Joint Research on E-Business Information Processing (2013B01035), National High Technology Research and Development Program of China (863 Program) (SS2014AA012303), Foundation for the Author of National Excellent Doctoral Dissertation of PR China (201189), and Program for New Century Excellent Talents in University (NCET-11-0778). J.Z. thanks the fund of the State Key Laboratory of Software Development Environment (SKLSDE-2015ZX-28).
\end{acknowledgments}


\end{document}